\documentclass[twocolumn,english,rmp]{revtex4}
\usepackage[T1]{fontenc}
\usepackage[latin9]{inputenc}
\usepackage{float}
\usepackage{amsmath}
\usepackage{graphicx}
\usepackage{amssymb}

\makeatletter

\usepackage{float}

\usepackage{textcomp}

\usepackage{esint}



\makeatother

\usepackage{babel}

\begin{document}

\title{Charged particle-display }

\author{Sung Nae Cho }

\email{sungnae.cho@samsung.com}

\affiliation{Micro Systems Laboratory, Samsung Advanced Institute of Technology,
Mt. 14-1 Nongseo-dong, Giheung-gu, Yongin-si, Gyeonggi-do 446-712,
South Korea }

\date{Revised 27 July 2009}

\begin{abstract}
An optical shutter based on charged particles is presented. The output
light intensity of the proposed device has an intrinsic dependence
on the interparticle spacing between charged particles, which can
be controlled by varying voltages applied to the control electrodes.
The interparticle spacing between charged particles can be varied
continuously and this opens up the possibility of particle based displays
with continuous grayscale. 
\end{abstract}


\maketitle

\section{Introduction }

The flexibility and bistability are the two key points in the next
generation display technologies. The flexibility implies the display
would be thin, lightweight, and ultimately, paper like, meaning it
would be cheap enough to be disposable. The bistability implies the
technology would be ecologically friendly. In the bistable display,
the image does not need to be refreshed until rewritten and, therefore,
a low level of power consumption is expected for still images. Unfortunately,
for motion pictures, the bistability has no advantages in power savings
over nonbistable technologies such as liquid crystal displays (LCDs),
plasma display panels (PDPs), and displays based on organic light
emitting diodes commonly referred to as OLEDs. 

The display technologies based on particles are the most prominent
candidates for a flexible and bistable displays. The earliest display
based on particles dates back to 1970's when Ota\cite{Ota} filed
for a patent. Since then various particle-displays based on electrophoresis
and electrowetting principles have emerged to form what is now referred
to as {}``E-paper technologies'' in the industry.\cite{twisting_ball,Mizuno,Matsuda,Jacobson,Ding}
In electrophoresis, particles are usually suspended in a fluid. Because
the speed at which particles move inversely vary with fluid density,
particle-displays based on electrophoresis have slow response time,
typically on the order of $300\,\textnormal{ms}.$\cite{nature,Whiteside,T. Kosc}
This makes motion pictures unsuitable for electrophoretic particle-displays.
The issue of slow response time in electrophoretic particle-displays,
however, has been resolved with the unveil of Quick Response Liquid
Powder Display (QR-LPD) by Hattori et al.\cite{QR_LPD1,QR-LPD2,QR-LPD3,QR_LPD4}
The QR-LPD is distinguished from the rest of particle based displays
in that it uses air as the particle carrying medium rather than fluid.
Because the particles in QR-LPD move in air, its response time is
at $0.2\,\textnormal{\textnormal{ms},}$ which is even faster than
LCDs. The submillisecond response time makes QR-LPD the only candidate
based on particle-display capable of handling motion pictures, and
researches are being conducted for particle-displays with air as the
particle carrying medium.\cite{LG2}

Common to all particle-displays, regardless of whether air or fluid
is used as the particle carrying medium, is the lack of continuous
grayscale. Here, the terminology, {}``continuous grayscale,'' is
referred to as the number of grayscale levels required to produce
desired number of colors. In principle, a display device with continuous
grayscale can generate an infinite range of colors. That being clarified,
a high grayscale range is essential to quality displays. Without it,
the images displayed on monitors would be dull.\cite{Stephen_Johnson}
Recently, it has been reported particle-display based on QR-LPD can
generate up to 16 grayscale levels (i.e., 4 bits), which corresponds
to the capability of generating 4096 colors.\cite{QR-LPD-16gray-level,QR-LPD-16gray-level-2}
To the proponents of other competing flat panel display technologies,
such as flexible LCDs, PDPs, or OLEDs, 4 bits of gray scale range
could hardly be considered a technological milestone. However, considering
only 4 grayscale levels were possible just a few years ago for particle-displays,
16 grayscale levels is a significant technological advancement for
the E-paper technology.\cite{QR-LPD-4gray-level}

The most well known E-paper technology, E-Ink, obtains different gray
states through modulation of voltages supplied to the control electrodes.
The same mechanism is employed by particle-displays based on QR-LPD
for achieving different gray states.\cite{T. Kosc,QR-LPD-4gray-level}
Because the voltage is modulated at a value lower than the saturation
voltage, which is the voltage required to display either all black
or all white for a simple black-and-white display, the displayed image
would have intensity somewhere between that of completely black and
completely white. This approach to achieve different gray states,
however, is done at the cost of losing bistability. Since the pixel
is being constantly modulated to sustain a gray state, the situation
is equivalent to motion pictures and the power savings from bistability
no longer applies to gray states. In spite of the lost bistability
advantages over the other competing flexible display technologies,
the speed at which particle can be modulated is limited by its finite
inertia (mass) and this places a practical limit on the extent to
which the grayscale levels of particle-display can be enhanced by
the aforementioned method. Most recently, Chim\cite{QR-LPD-64gray-level}
demonstrated a 64 grayscale levels (or 6bits) for particle-display
based on QR-LPD for his masters thesis project at Delft University
of Technology. However, to generate 64 grayscale levels, the DATA
Driver, which is a serial to parallel shift register that shifts data
words of 6 bits per clock cycle, is required.

\begin{figure}[H]
\begin{centering}
\includegraphics[scale=0.4]{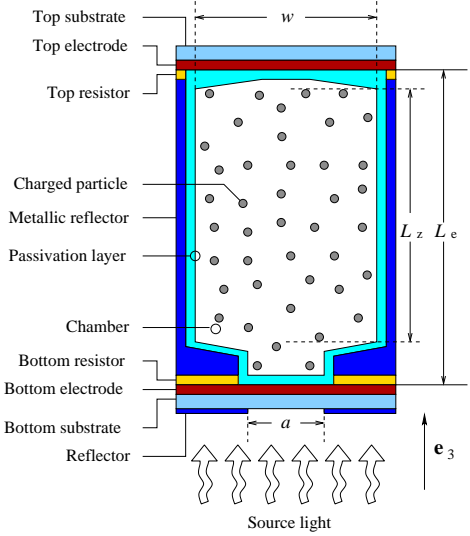} 
\par\end{centering}

\caption{\label{fig:transflective_0} (Color online) Transflective display
based on charged particles. }

\end{figure}

In this work, a display architecture based on charged particles is
presented. Unlike the previous particle-display technologies, the
proposed device has potential to generate continuous grayscale levels
without sacrificing the benefit of bistability. Also,  the proposed
particle-display technology, in principle, does not require complicated
voltage modulation schemes to generate different gray states.\cite{SNCHO}

\section{Charged particle-display}

\subsection{Device structure}

The cross-sectional schematic of an optical shutter based on charged
particles of same polarity is illustrated in Fig. \ref{fig:transflective_0}.
In the figure, the two electrodes, where each is labeled top and bottom
electrodes, constitute the control electrodes. Because the light must
be transmitted through the control electrodes, the electrodes are
chosen from optically transparent conductors. The chamber, wherein
the particles reside, can be a vacuum, filled with noble gas, or filled
with air. The metallic reflectors, which forms the lateral surface
of the chamber, are electrically connected to one of the control electrodes.
This make metallic reflectors not only to reflect light, but also
function as to keep particles from aggregating to the lateral surface
of chamber. The optically transparent passivation layer, which is
treated on the inner surface of the chamber, functions to prevent
charge transfer between particles and conductors. For positively charged
particles inside the chamber, the hydrophobic treatment on the surface
of the passivation layer, which is not explicitly shown in the figure,
for example, prevents particles from sticking to the surface.

\begin{figure}[H]
\begin{centering}
\includegraphics[scale=0.4]{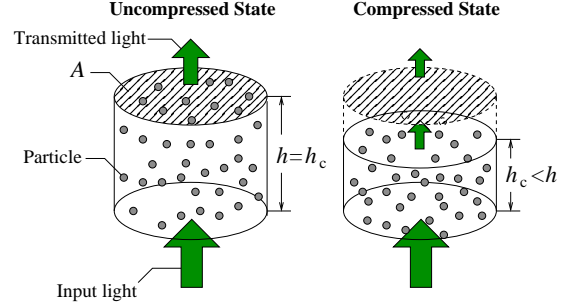} 
\par\end{centering}

\caption{\label{fig:cylinder_0} (Color online) Transmission of light through
a medium filled with total of $N$ suspended spherical particles.
Both of the compressed and the uncompressed cases are shown. }

\end{figure}

\subsection{Operation principle}

The intensity of light transmitted through a medium filled with suspended
spherical particles, illustrated in Fig. \ref{fig:cylinder_0}, is
given by \begin{equation}
I=I_{0}\exp\left[-\frac{36n\pi kfh}{\lambda\left(n^{2}+2\right)^{2}}\right],\label{eq:transmitted_intensity_BH}\end{equation}
 where $n$ and $k$ are, respectively, the real and the imaginary
part of the complex refractive index for the medium and the particles
suspended in it; $I_{0}$ is the initial intensity, $\lambda$ is
the wavelength of incidence light, $h$ is the height (or thickness)
of the volume containing charged particles, and $f$ is the volume
fraction of particles.\cite{Bohren_Huffman} In explicit form, the
volume fraction of particles is \[
f=\frac{Nv_{\textup{p}}}{Ah_{\textup{c}}},\]
 where $N$ is the total number of particles in the optical chamber,
$v_{\textup{p}}$ is the volume of a single particle, $h_{\textup{c}}$
is the height of compressed cylinder containing charged particles,
and $A$ is the cross-sectional area of the cylindrical chamber illustrated
in Fig. \ref{fig:cylinder_0}. Throughout this work, I shall refer
to terms such as {}``compressed state'' or {}``compressed particle
volume'' to denote the compression of the volume containing particles.
Insertion of $f$ into Eq. (\ref{eq:transmitted_intensity_BH}) gives
\begin{eqnarray}
I=I_{0}\exp\left(-\frac{C}{h_{\textup{c}}}\right), &  & C=\frac{36n\pi khNv_{\textup{p}}}{\lambda\left(n^{2}+2\right)^{2}A},\label{eq:transmitted_intensity}\end{eqnarray}
 where $C$ is just a constant.

The complex dielectric constant, $\tilde{\varepsilon},$ and the complex
refractive index, $\tilde{n},$ are expressed in form as \begin{eqnarray}
\tilde{\varepsilon}=\varepsilon_{1}+i\varepsilon_{2}, &  & \tilde{n}=n+ik,\label{eq:complex-optical-constants}\end{eqnarray}
 where $\varepsilon_{1}$ and $n$ are the real parts, and $\varepsilon_{2}$
and $k$ are the imaginary parts. For the complex refractive index,
$n$ represents the real refractive index and $k$ is the extinction
coefficient (or the absorption coefficient). The two, $\tilde{\varepsilon}$
and $\tilde{n},$ are related by the expression \begin{align*}
\tilde{\varepsilon}-\tilde{n}^{2} & =0.\end{align*}
 With the real and imaginary parts inserted for $\tilde{\varepsilon}$
and $\tilde{n}$ from Eq. (\ref{eq:complex-optical-constants}), I
have \begin{align*}
\varepsilon_{1}+i\varepsilon_{2}-\left(n+ik\right)^{2} & =0.\end{align*}
 The resulting expression can be rearranged to yield \begin{align}
\left(\varepsilon_{1}-n^{2}+k^{2}\right)+i\left(\varepsilon_{2}-2nk\right) & =0.\label{eq:complex-optical-constants-algebra}\end{align}
 Equation (\ref{eq:complex-optical-constants-algebra}) can only be
true if and only if the real and the imaginary parts are equal to
zero independently. This requirement yields the two expressions connecting
$\left(\varepsilon_{1},\varepsilon_{2}\right)$ to the optical constants
$\left(n,k\right);$ and, the two expressions are \begin{eqnarray}
\varepsilon_{1}=n^{2}-k^{2}, &  & \varepsilon_{2}=2nk.\label{eq:e1-e2-complex-dielectric-constant}\end{eqnarray}

Stoller et al.\cite{gold-particle-quantum} measured the complex dielectric
constant, $\tilde{\varepsilon}\left(\varepsilon_{1},\varepsilon_{2}\right),$
for the single gold nanoparticle. For the gold nanoparticles of diameters
$10$ and $15\,\textup{nm},$ assuming a spherical morphology, they
observed a reasonably good correspondence existing between complex
dielectric constants of the bulk gold and the gold nanoparticles for
the wavelength range of roughly $510$ to $580\,\textup{nm}.$ Their
finding is of significant importance, as it allows the bulk gold dielectric
constants, which is readily available, to be used for the gold nanoparticles,
which is not so readily available. Since the real and the imaginary
parts of the complex dielectric constant are related to the optical
constants $\left(n,k\right)$ thru Eq. (\ref{eq:e1-e2-complex-dielectric-constant}),
the $n$ and $k$ for the gold nanoparticles are readily available
once $\varepsilon_{1}$ and $\varepsilon_{2}$ are known from the
bulk counterpart. The vice versa is also true, of course. Fortunately,
Johnson and Christy measured the optical constants for the bulk copper,
silver, and the bulk gold.\cite{gold-bulk} Justified by the findings
of Stoller et al.,\cite{gold-particle-quantum} for $\lambda=550\,\textup{nm},$
assuming the gold nanoparticle of radius $r_{\textup{p}}=7.5\,\textup{nm},$
the $n$ and $k$ measurements from the work of Johnson and Christy,\cite{gold-bulk}
\begin{eqnarray}
\textup{bulk gold:}\;\tilde{n}=n+ik, &  & \left\{ \begin{array}{c}
n=0.43,\quad\:\\
k=2.455,\;\:\:\\
\lambda=550\,\textup{nm},\end{array}\right.\label{eq:bulk-gold-n-k}\end{eqnarray}
 can be utilized for the gold nanoparticles to plot Eq. (\ref{eq:transmitted_intensity})
for the transmitted intensity.

For a spherical gold nanoparticle, the particle volume, $v_{\textup{p}},$
is given by \begin{align*}
v_{\textup{p}} & =\frac{4}{3}\pi r_{\textup{p}}^{3},\end{align*}
 where $r_{\textup{p}}$ is the particle radius. Assuming the diameter
of $15\,\textup{nm}$ for the gold nanoparticle, $v_{\textup{p}}$
becomes \begin{eqnarray}
r_{\textup{p}}=7.5\,\textup{nm}, &  & v_{\textup{p}}=1.77\times10^{-24}\,\textup{m}^{3}.\label{eq:vp}\end{eqnarray}
 To plot Eq. (\ref{eq:transmitted_intensity}), I shall assume the
following values for the chamber parameters, (the height $h$ and
the cross-sectional area $A$), and the particle number, $N,$ \begin{equation}
\left\{ \begin{array}{c}
h=100\,\mu\textup{m}=1\times10^{-4}\,\textup{m},\,\\
A=\pi h^{2}=3.14\times10^{-8}\,\textup{m}^{2},\\
N=2.46\times10^{8}.\quad\quad\quad\qquad\,\end{array}\right.\label{eq:hAN}\end{equation}

\begin{figure}[H]
\begin{centering}
\includegraphics[scale=0.4]{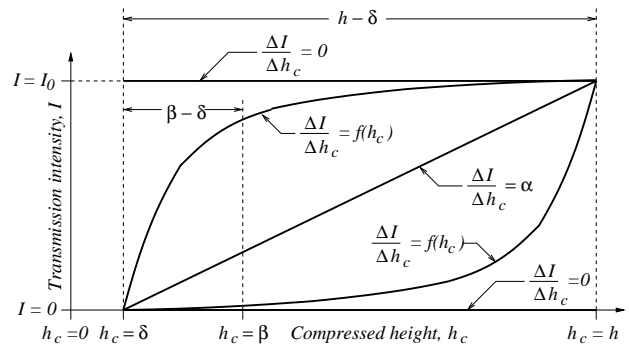} 
\par\end{centering}

\caption{\label{fig:transmission_intensity_schematic} Schematic of curves
illustrating the intensity, $I,$ of transmitted light as a function
of compression height, $h_{\textup{c}}.$ }

\end{figure}

Before going ahead with plotting Eq. (\ref{eq:transmitted_intensity}),
it is worthwhile to double check if the total number of particles,
$N,$ specified in Eq. (\ref{eq:hAN}) is reasonable. With chamber
parameters as defined in Eq. (\ref{eq:hAN}), the volume of the cylindrical
chamber is \[
V=Ah=\pi h^{3}.\]
 Suppose the charged spherical particles can be compressed and enclosed
in a confined volume in such a way that neighboring particles actually
touch each other. How many particles, under such restrictions, can
be fitted inside the chamber defined by Eq. (\ref{eq:hAN})? The answer
is $N_{\textup{max}}$ and its expression is \begin{equation}
N_{\textup{max}}=\frac{V}{v_{\textup{p}}}=\frac{3}{4}\left(\frac{h}{r_{\textup{p}}}\right)^{3}.\label{eq:Nmax_particle_num}\end{equation}
 For the spherical particle of radius $r_{\textup{p}}=7.5\,\textup{nm},$
the $N_{\textup{max}}$ is roughly \begin{equation}
N_{\textup{max}}\approx1.7\times10^{12}.\label{eq:Nmax_number}\end{equation}
 For the charged particles, the condition of neighboring particles
actually touching each other is not possible due to Coulomb repulsion,
unless the external compression force is infinite. Nonetheless, the
expression for $N_{\textup{max}},$ defined in Eq. (\ref{eq:Nmax_particle_num}),
provides the upper limit for $N$ inside the chamber.

The physical optical shutter based on charged particles must be compressible
in order to allow variations in transmission intensity, which imposes
the condition, \[
N<N_{\textup{max}}.\]
 Equivalently, the variability of transmission intensity thru compression
requires $N\neq0$ and this further modifies the condition for $N$
as \begin{equation}
0<N<N_{\textup{max}}.\label{eq:zero_N_Nmax}\end{equation}
 The inequality condition for $N,$ defined in Eq. (\ref{eq:zero_N_Nmax}),
can be understood from the illustration shown in Fig. \ref{fig:transmission_intensity_schematic},
where some of the possible curves for the transmission intensity,
$I,$ as a function of the compression height, $h_{\textup{c}},$
are shown. The cases where $N=0$ and $N=N_{\textup{max}}$ are represented
by the curves corresponding to $\triangle I/\triangle h_{\textup{c}}=0$
at $I=I_{0}$ and $I=0,$ respectively. Similarly, the the curve corresponding
to $\triangle I/\triangle h_{\textup{c}}=0$ at $I=0$ in the figure
represents the case where $N=N_{\textup{max}}.$ For $N=N_{\textup{max}},$
the particles in the chamber are already maximally compressed and,
therefore, no light gets transmitted. On the other hand, for $N=0,$
all light gets transmitted through the optical shutter, as the inside
of the optical shutter is a void.

\begin{figure}[H]
\begin{centering}
\includegraphics[scale=0.4]{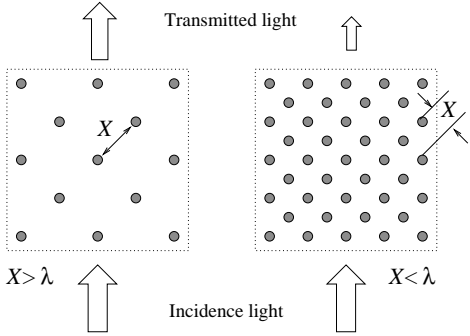} 
\par\end{centering}

\caption{\label{fig:grid} Array of metallic nanoparticles in grid formation.
The parameter $X$ is the grid (or particle-particle) spacing and
$\lambda$ is the wavelength of the incidence light. For the charged
particles inside the chamber, the parameter $X$ can only be thought
of as the time averaged mean particle-particle distance between the
nearest neighboring particles. Nonetheless, the grid representation
illustrated here serves the purpose of initiating the crude connection
between the wavelength of incidence light, $\lambda,$ and the total
particle number, $N.$ }

\end{figure}

For the case of any finite $N$ satisfying the condition defined in
Eq. (\ref{eq:zero_N_Nmax}), the curve for the transmission intensity
must necessarily lie in the region between the two extreme cases,
$N=0$ and $N=N_{\textup{max}},$ as schematically demonstrated in
Fig. \ref{fig:transmission_intensity_schematic}. Three such curves,
corresponding to finite $N,$ are illustrated in the figure: (1) the
upper curve represented by $\triangle I/\triangle h_{\textup{c}}=f\left(h_{\textup{c}}\right),$
(2) the curve represented by $\triangle I/\triangle h_{\textup{c}}=\alpha,$
and (3) the lower curve represented by $\triangle I/\triangle h_{\textup{c}}=f\left(h_{\textup{c}}\right),$
where $\alpha$ is a constant and $f\left(h_{\textup{c}}\right)$
is a function of $h_{\textup{c}}.$ The two curves corresponding to
$\triangle I/\triangle h_{\textup{c}}=f\left(h_{\textup{c}}\right)$
only appreciably varies in transmission intensity with compression
within the window of $\beta-\delta,$ where $\beta-\delta\ll h.$
Such curves are associated with $N,$ in which the $N$ is finite
but close to either $N=0$ or $N=N_{\textup{max}}.$ Because the transmission
intensity only appreciably varies within $\beta-\delta\ll h$ for
such choices of $N,$ the gray states are difficult to achieve as
it requires very precise control of the compression height. For example,
assuming the particles can be compressed at the increment in $\triangle h_{\textup{c}},$
the fact that $\beta-\delta\approx\triangle h_{\textup{c}}$ makes
it difficult to achieve gray states. To achieve gray states, the $\triangle h_{\textup{c}},$
which defines the sensitivity of compression, must be much smaller
than $\beta-\delta.$ Consequently, the transmission intensity curves
for the aforementioned choices of $N$ are only good for displaying
either completely bright or completely dark transmission states.

\begin{figure}[H]
\begin{centering}
\includegraphics[scale=0.4]{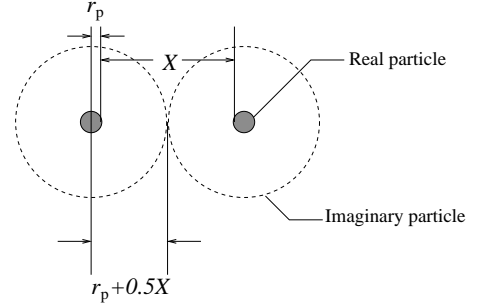} 
\par\end{centering}

\caption{\label{fig:particle_num} The nanoparticles of radius $r_{\textup{p}}$
are separated by $X,$ which is the nearest surface to surface separation
between two neighboring particles. The imaginary particles of radius
$r_{\textup{I}}=r_{\textup{p}}+0.5X$ are, however, in closed packed
formation for a given nearest surface to surface separation of $X.$ }

\end{figure}

The optimal design for the presented optical shutter is achieved by
choosing $N$ for the total particle number in the chamber in such
way that the curve for the transmission intensity goes like $\triangle I/\triangle h_{\textup{c}}=\alpha,$
where $\alpha$ is a constant, in Fig. \ref{fig:transmission_intensity_schematic}.
Because the curve is linear for the transmission intensity, the window
of range in which compression can be done is maximized, $h-\delta\lesssim h.$
For this particular choice of $N,$ the $\triangle h_{\textup{c}},$
which is the increment at which particles can be compressed, is much
less than $h-\delta,$ i.e., $h-\delta\gg\triangle h_{\textup{c}},$
and the number of different gray states that can be achieved is given
by \begin{align}
N_{\textup{gray}} & =\frac{h-\delta}{\triangle h_{\textup{c}}}.\label{eq:Ngray}\end{align}

In principle, the $\triangle h_{\textup{c}}$ can be made finer and
finer as desired. In reality, the fineness of $\triangle h_{\textup{c}}$
is limited by the system design. Nevertheless, with the right choice
of $N,$ the grayscale levels in number of $N_{\textup{gray}},$ as
defined in Eq. (\ref{eq:Ngray}), is possible with the presented optical
shutter based on charged particles. The question remains to be answered
is this: how do we go about obtaining the right $N$? To answer this,
I shall refer to the illustration shown in Fig. \ref{fig:grid}.

Due to the wave nature of light, the total particle number and the
transmission intensity depend on the wavelength, $\lambda,$ of the
incidence light. As a rudimentary assumption, the transmission loss
of an electromagnetic wave passing through an array of metallic particles
decreases for $X\gg\lambda$ and increases for $X\lesssim\lambda/j,$
where $j=2,\,3,\,\cdots,$ which is schematically illustrated in Fig.
\ref{fig:grid}. The case where $X\gg\lambda$ represents the situation
in which $N$ is very small, whereas the case where $X\lesssim\lambda/j,$
represents the situation in which $0\ll N\ll N_{\textup{max}},$ provided
the $j$ is not too large. It is, therefore, not too bad to impose
the condition, $X\approx\lambda/2,$ for $X$ in estimating for the
total number of particles in the chamber. For $\lambda=550\,\textup{nm},$
this condition for $X$ yields the value of $X=275\,\textup{nm}.$
To figure out exactly how man particles can be fitted inside the chamber
under restriction $X\approx\lambda/2,$ the Fig. \ref{fig:particle_num}
is referred to. Since $X$ is the closest distance between surfaces
of two nearest neighbor particles, I shall visualize an imaginary
particle of radius $r_{\textup{I}},$ \begin{align}
r_{\textup{I}} & =r_{\textup{p}}+0.5X.\label{eq:r_imaginary}\end{align}
 Assuming the imaginary particles are close-packed inside the chamber,
the problem becomes identical to the previous case, which resulted
in Eq. (\ref{eq:Nmax_particle_num}). With $r_{\textup{I}}$ of Eq.
(\ref{eq:r_imaginary}) inserted for $r_{\textup{p}}$ in Eq. (\ref{eq:Nmax_particle_num}),
the expression becomes \begin{eqnarray}
N_{\textup{min}}=\frac{3}{4}\left(\frac{h}{r_{\textup{p}}+0.5X}\right)^{3}, &  & X=\frac{\lambda}{2},\label{eq:Nmim_particle_num}\end{eqnarray}
 where $N_{\textup{max}}$ has been replaced by $N_{\textup{min}}.$
With $N_{\textup{min}}$ defined in Eq. (\ref{eq:Nmim_particle_num}),
the number of particles inside the chamber may be chosen according
to the inequality, \begin{equation}
0\ll N_{\textup{min}}\lesssim N\ll N_{\textup{max}}.\label{eq:N_condition}\end{equation}
 It can be easily verified that $N_{\textup{min}}\gg0.$ For $X=275\,\textup{nm},$
which corresponds to the half wavelength of $\lambda=550\,\textup{nm},$
$N_{\textup{min}}$ is roughly $N_{\textup{min}}\approx2.46\times10^{8}.$
This value for $N_{\textup{min}}$ is much larger than zero, but it
is much smaller than $N_{\textup{max}},$ which has the value $N_{\textup{max}}\approx1.7\times10^{12}$
from Eq. (\ref{eq:Nmax_number}). For the optical shutter involving
charged particles of spherical morphology and the cylindrical chamber
of specifications  defined in Eq. (\ref{eq:hAN}), the inequality
condition for $N,$ Eq. (\ref{eq:N_condition}), becomes \begin{eqnarray}
\frac{3}{4}\left(\frac{h}{r_{\textup{p}}+\frac{\lambda}{2n}}\right)^{3}\lesssim N\ll\frac{3}{4}\left(\frac{h}{r_{\textup{p}}}\right)^{3}, &  & n=2,3,\cdots.\label{eq:N_condition_specific}\end{eqnarray}

\begin{figure}[H]
\begin{centering}
\includegraphics[scale=0.7]{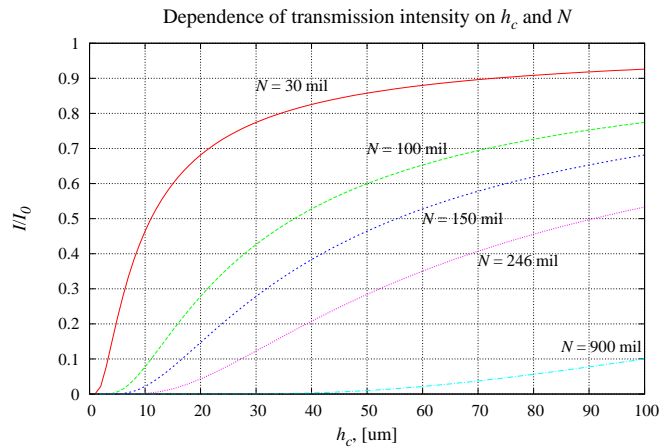} 
\par\end{centering}

\caption{\label{fig:transmission_intensity} (Color online) Transmission intensity
as a function of particle number, $N,$ and the compression height,
$h_{\textup{c}}.$ The abbreviation {}``mil'' in the figure denotes
a million, i.e., $1\times10^{6}.$ }

\end{figure}

Equation (\ref{eq:transmitted_intensity}) has been computed using
gold nanoparticles as the charged particles (the gold nanoparticle
was chosen only because its optical constant data, $n$ and $k,$
were readily available). For the charged particles and the incidence
light, the parameters defined in Eq. (\ref{eq:bulk-gold-n-k}) were
used. The parameters defined in Eq. (\ref{eq:hAN}) were used for
the chamber specification. Using Eq. (\ref{eq:transmitted_intensity}),
the transmission intensity for different values of $N$ were considered
and the results are shown in Fig. \ref{fig:transmission_intensity}.
The $N=30\times10^{6}$ curve is the case where particle number is
relatively low in the chamber. As it can be observed, for low particle
numbers in the chamber, the intensity does not vary well with compression
except for small $h_{\textup{c}},$ which is consistent with the curve
represented by $\triangle I/\triangle h_{\textup{c}}=f\left(h_{\textup{c}}\right)$
in Fig. \ref{fig:transmission_intensity_schematic}. Contrarily, the
$N=900\times10^{6}$ curve corresponds to the case where too many
particles are inside the chamber. Although the transmission intensity
varies linearly with compression, which is a good characteristic of
an optical shutter, the output intensity is far too low even for the
brightest state. For $N=900$$\times10^{6},$ the brightest state
only transmits $10\%$ of the initial input intensity, i.e., $I/I_{0}=0.1.$
As expected, the $N=246$$\times10^{6}$ curve in Fig. \ref{fig:transmission_intensity},
which corresponds to the case where $X=\lambda/2$ for $\lambda=500\,\textup{nm},$
most resembles the curve represented by $\triangle I/\triangle h_{\textup{c}}=\alpha$
in Fig. \ref{fig:transmission_intensity_schematic}. However, the
brightest output intensity is only $\sim50\%$ of the input intensity
of the incidence light. Compared to LCDs, where only $\sim5\%$ of
the intensity of the incidence light from back light unit gets transmitted,
the output intensity of $\sim50\%$ is already $10$ times more efficient
than LCDs.

\begin{figure}[H]
\begin{centering}
\includegraphics[scale=0.5]{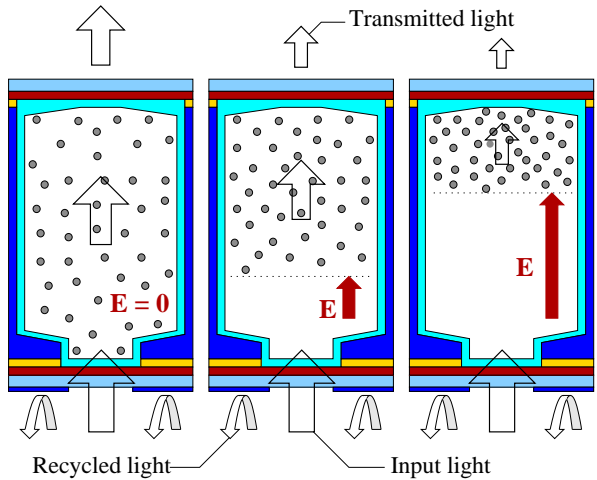} 
\par\end{centering}

\caption{\label{fig:transflective_1} (Color online) The on set of electric
field compresses the filled volume for charged particles.}

\end{figure}

The same principle, which is inherent in Eq. (\ref{eq:transmitted_intensity}),
applies to the proposed optical shutter, Fig. \ref{fig:transflective_0}.
In the presented device, the chamber is filled with charged particles
of same polarity and this can be identified with the medium filled
with suspended particle in Fig. \ref{fig:cylinder_0}. The on set
of electric field inside the chamber, which is done by controlling
the voltage over one of the electrodes, causes particles to be compressed
in volume, as illustrated in Fig. \ref{fig:transflective_1}. Since
the particles are assumed to be positively charged, they are compressed
in the direction of electric field. Eventually, the compression comes
to a stop when particle-particle Coulomb repulsion counterbalances
the compression induced by the control electrodes.

In principle, the level of compression for the particle volume can
be varied continuously. Because the intensity of transmitted light
varies with compression, i.e., Eq. (\ref{eq:transmitted_intensity}),
the display based on charged particles has the potential to generate
continuous gray levels. Also, since the control electrodes form a
capacitor, provided there is no leakage (or negligible) current across
the capacitor, the electric field inside the chamber can be sustained
even when the device is disconnected from power. This opens up the
possibility of a bistable mode for gray states as well.

\begin{figure}[H]
\begin{centering}
\includegraphics[scale=0.45]{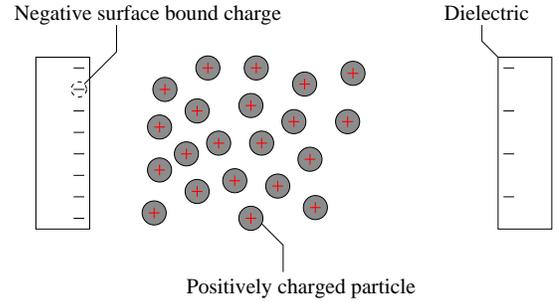} 
\par\end{centering}

\caption{\label{fig:BC1} (Color online) Charged particles near dielectric
wall.}

\end{figure}

\begin{figure}[H]
\begin{centering}
\includegraphics[scale=0.45]{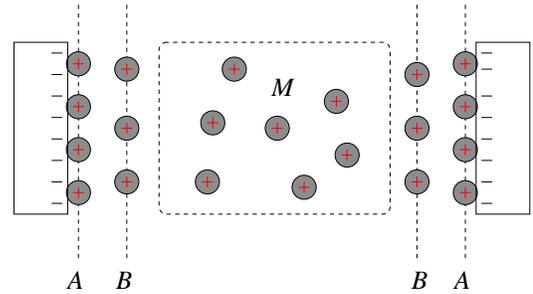} 
\par\end{centering}

\caption{\label{fig:BC2} (Color online) Charged particles in region $M$ are
shielded from the negatively charged surface bound charges within
the dielectric wall. }

\end{figure}

The presented device works as an optical shutter, provided the charged
particles can be effectively prevented from piling up at the surface
of dielectric walls. When a charged particle is brought close to the
dielectric surface, the bound charges within the dielectric get redistributed
in order to reduce the field originating from the charged particle
placed near vicinity of dielectric surface. Such is illustrated in
Fig. \ref{fig:BC1}. For the case where a positively charged particle
is placed near the surface of a dielectric, the surface bound charges
of opposite polarity get induced and distributed near the inner surface
of dielectric. As a result, the positively charged external particle
gets pulled to the dielectric surface and, eventually, sticking to
the surface of dielectric, which process is illustrated in Fig. \ref{fig:BC2}.
If there are more than one positively charged particles placed at
the vicinity of dielectric surface, the aforementioned process continues
and, eventually, layers get formed, for example, the layers $A$ and
$B$ in the figure. This process does not continue indefinitely, however,
as the bound charges of opposite polarity within the dielectric get
eventually shielded by the positively charged particles forming layers
over the dielectric surface. Because the particles in each layer are
charged with same polarity, the Coulomb repulsion keeps the two particles
from touching each other. Assuming the layer $B$ is sufficient to
shield completely the negatively charged surface bound charges within
the dielectric, the remaining positively charged particles in the
region between two dielectric walls would have no other places to
go, except to continuously bounce back and forth within the region,
which region has been indicated by $M$ in Fig. \ref{fig:BC2}. 

The aforementioned description for an optical shutter, however, has
a serious problem which is associated with the particle layers forming
on the surface of dielectric walls. Assuming the path of light propagation
is along the horizontal axis, the light enters the device from the
left and exits at the right or vice versa. In principle, the compression
of charged particles in region, $M,$ controls the intensity of transmitted
light. The problem arises because the charged particles forming layers
$A$ and $B$ on the surface of dielectric walls may no longer be
optically transparent. As previously discussed, using the illustration
in Fig. \ref{fig:grid} as an example, metallic particles in an array
of grid formation severely reduces the intensity of transmitted light
for grid spacing, $X,$ much less than the wavelength, $\lambda,$
of the incidence light. I now discuss the ways to resolve this complication. 

\begin{figure}[H]
\begin{centering}
\includegraphics[scale=0.45]{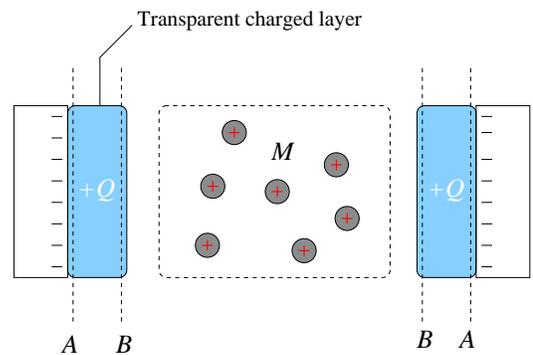} 
\par\end{centering}

\caption{\label{fig:BC3} (Color online) Transparent charged layer formed by
ISAM or electron beam irradiation method effectively plays the role
of positively charged particle layers at $A$ and $B.$ }

\end{figure}

It is well known that the surface of a typical $SiO_{2}$ glass becomes
hydrophilic in an open air as a result of the silane group at the
surface of glass combining with oxygen.\cite{hydrophilic_1,hydrophilic_2}
This has the effect of making the surface of glass to be slightly
negatively charged and any positively charged particles in vicinity
would be attracted to the glass surface forming layers such as those,
e.g., $A$ and $B,$ illustrated in Fig. \ref{fig:BC2}. Unless the
attracted, positively charged, particles are optically transparent,
the device would not be able to function as an optical shutter for
no light would to pass through the device. Necessarily, the role of
layers $A$ and $B$ in Fig. \ref{fig:BC2} must be played out by
particles of same polarity as those residing in region $M,$ but distinguished
from those residing in region $M$ in that they are optically transparent.
One way to achieve this is to chemically treat the surface of glass
so as to make it hydrophobic. The hydrophobic treatment of glass electrically
neutralizes the glass surface, thereby significantly reducing the
oppositely charged particles from sticking to the glass surface. For
large and weakly charged positive particles, simple electrical neutralization
of glass surface by hydrophobic treatment is sufficient to keep particles
from permanently sticking to the surface. In such system, charged
particles are continuously bounced off of the wall and do not stick
to the surface. However, for smaller and well charged positive particles,
simple hydrophobic treatment of glass surface is not sufficient to
prevent oppositely charged particles from sticking to the glass surface.
For such cases, it is necessary to make the surface of glass net positively
charged. This can be easily done by utilizing the technique known
as {}``ionically self assembled monolayer'' (ISAM), in which technique
charged particles, polymers, or monomers of preferred polarity are
physically attached to the surface.\cite{charged_glass_4_ISAM} Alternatively,
and more directly, the glass may be irradiated with electron beam
to physically embed charged particles inside the glass medium.\cite{charged_glass_1,charged_glass_2}
In this latter method, for example, net negative charges may be made
to accumulate inside the glass. This has the effect of inducing positive
charges on the surface of glass, which repels positive charged particles
in vicinity of the glass surface, thereby preventing particles from
sticking to the surface. In summary, the role of positively charged
particle layers, $A$ and $B$ in Fig. \ref{fig:BC2}, may be effectively
get taken cared by the aforementioned ISAM or the electron beam irradiation
methods; and, using these alternative techniques, the glass surface
repels charged particles inside the chamber and it is optically transparent,
as illustrated in Fig. \ref{fig:BC3}. 

To utilize charged particles in displays, a quantitative understanding
of how design parameters, such as $\left(L_{\textup{e}},L_{\textup{z}},\xi\right)$
for the sub-pixel dimensions and $\left(r,\rho_{\textup{m}},Q\right)$
for the charged particle, enter into the compression mechanism is
required. Here, $r$ is the particle radius (assuming a spherical
particle), $\rho_{\textup{m}}$ is the particle mass density, and
$Q$ is the net charge which the particle holds. Describing the compression
mechanism in terms of the aforementioned design parameters is the
task for the next section.

\subsection{Theory}

The optical shutter presented in this proposal relies on the density
of particles in chamber to control the intensity of transmitted light.
An analogy can be made to the driving under misty weather. When the
density of water vapor suspended in the atmosphere is heavy, one is
obscured in his or her viewing distance, as less light reaches the
eye. Contrarily, the amount of light reaching the eye increases with
a reduction in density of water vapor suspended in the atmosphere,
thereby enabling the driver to see far distances.

The particles in chamber of the proposed optical shutter ranges in
diameter anywhere from a few nanometers to several microns, assuming
a spherical morphology. This range for the particle size, although
small macroscopically, is much too large to be considered for a treatment
within quantum domain, where the quantum theory must be used for a
description. Therefore, the classical theory suffices for the description
here. Since the charged particles in the system, as a whole, behave
like a classical gas, the description is carried out in the realm
of statistical physics.

To keep the topic presented here self-contained, I shall briefly summarize
the kind of manipulations and approximations assumed in obtaining
expressions which are considered crucial to the initial development
of the analysis.

\subsubsection{Maxwell-Boltzmann statistics}

The charged particles in chamber can be treated as classical particles
obeying the Maxwell-Boltzmann statistics.\cite{Reif} The $i\textup{th}$
charged particle under influence of external forces, for example,
gravitational and electric forces, assumes the energy \begin{eqnarray*}
U_{i}=\frac{\mathbf{p}_{i}^{2}}{2m_{i}}+U_{\textup{ext}}+U_{\textup{int}}, &  & \mathbf{p}_{i}=\sum_{j}p_{ij}\mathbf{e}_{j},\end{eqnarray*}
 where $\mathbf{p}_{i}$ is the center of mass momentum for the $i\textup{th}$
charged particle and the term associated with it is the kinetic energy,
$U_{\textup{ext}}$ is the interaction energy with external influences,
and $U_{\textup{int}}$ is the energy contribution arising only if
the particle is not monatomic.

In explicit form, $U_{\textup{ext}}$ can be expressed as \begin{align*}
U_{\textup{ext}}= & \sum_{j\neq i}^{N}\frac{k_{\textup{q}}Q_{i}Q_{j}}{\left[\left(x_{i}-x_{j}\right)^{2}+\left(y_{i}-y_{j}\right)^{2}+\left(z_{i}-z_{j}\right)^{2}\right]^{1/2}}+m_{i}gz_{i}\\
 & +Q_{i}Ez_{i},\end{align*}
 where $N$ is the number of particles in volume, the $Q_{i}$ and
$Q_{j}$ denote respectively the net charges for the $i\textup{th}$
and $j\textup{th}$ particles, $E$ is the electric field magnitude,
and the constants $g=9.8\,\textup{m}\,\textup{s}^{-2}$ and $k_{\textup{q}}=8.99\times10^{9}\textup{ N}\,\textup{m}^{2}\,\textup{C}^{-2}$
in MKS system of units. For a monatomic particle, the energy contributions
from the internal rotation and vibration with respect to its center
of mass vanishes, $U_{\textup{int}}=0.$ Therefore, the $i\textup{th}$
monatomic charged particle under the influence of external forces
assumes the energy \begin{align}
U_{i}= & \frac{\mathbf{p}_{i}^{2}}{2m_{i}}+m_{i}gz_{i}+QEz_{i}\nonumber \\
 & +\sum_{j\neq i}^{N}\frac{k_{\textup{q}}Q^{2}}{\left[\left(x_{i}-x_{j}\right)^{2}+\left(y_{i}-y_{j}\right)^{2}+\left(z_{i}-z_{j}\right)^{2}\right]^{1/2}},\label{eq:Ui}\end{align}
 where, for convenience, all particles in the system are assumed to
be identically charged with same polarity, i.e., $Q_{i}=Q_{j}=Q.$

The gravity and electric field have directions, and this information
must be taken into account in Eq. (\ref{eq:Ui}). To do this, the
parameter $z_{i}$ is first restricted to a domain \[
\left\{ \mathcal{D}:\;0\leq z_{i}<\infty\right\} .\]
 With $z_{i}$ restricted to a domain defined by $\mathcal{D},$ the
gravitational potential energy of a particle, $m_{i}gz_{i},$ increases
with positive $g$ and decreases with negative $g$ as $z_{i}$ increase.
Therefore, the direction of gravity in Eq. (\ref{eq:Ui}) can be taken
into account by \begin{equation}
g=\left\{ \begin{array}{c}
+9.8\,\textup{m}\,\textup{s}^{-2}\,\textup{if gravity points in }-\mathbf{e}_{3},\\
-9.8\,\textup{m}\,\textup{s}^{-2}\,\textup{if gravity points in }+\mathbf{e}_{3}.\end{array}\right.\label{eq:g_direction}\end{equation}
 The electric potential energy of a particle, $QEz_{i},$ increases
with positive $E$ and decreases with negative $E$ as $z_{i}$ increase.
For a positively charged particle, its electric potential energy increases
as it moves against the direction of electric field and decreases
as it moves in the direction of electric field. The direction of electric
field inside the chamber can hence be taken into account in Eq. (\ref{eq:Ui})
by \begin{equation}
E=\left\{ \begin{array}{c}
+E\,\textup{if }\mathbf{E}\,\textup{points in }-\mathbf{e}_{3},\\
-E\,\textup{if }\mathbf{E}\,\textup{points in}+\mathbf{e}_{3}.\end{array}\right.\label{eq:E_direction}\end{equation}

The probability of finding the particle with its center of mass position
in the ranges $\left(\mathbf{R}_{i};d\mathbf{R}_{i}\right)$ and $\left(\mathbf{p}_{i};d\mathbf{p}_{i}\right)$
can be expressed as \begin{equation}
P_{s}\left(\mathbf{R}_{i},\mathbf{p}_{i}\right)d^{3}\mathbf{R}_{i}d^{3}\mathbf{p}_{i}\propto\exp\left(-\frac{U_{i}}{k_{\textup{B}}T}\right)d^{3}\mathbf{R}_{i}d^{3}\mathbf{p}_{i},\label{eq:Ps0}\end{equation}
 where $T$ is the temperature in units of degree Kelvin $\left(^{o}\textup{K}\right)$
and $k_{\textup{B}}$ is the Boltzmann constant, $k_{\textup{B}}=1.38\times10^{-23}\textup{J}/\left(^{o}\textup{K}\right).$
With Eq. (\ref{eq:Ui}), $P_{s}\left(\mathbf{R}_{i},\mathbf{p}_{i}\right)d^{3}\mathbf{R}_{i}d^{3}\mathbf{p}_{i}$
becomes \begin{align}
 & P_{s}\left(\mathbf{R}_{i},\mathbf{p}_{i}\right)d^{3}\mathbf{R}_{i}d^{3}\mathbf{p}_{i}\nonumber \\
 & \propto\exp\left(-\frac{\mathbf{p}_{i}^{2}}{2m_{i}k_{\textup{B}}T}\right)d^{3}\mathbf{p}_{i}\nonumber \\
 & \times\exp\left(\sum_{j\neq i}^{N}\frac{-k_{\textup{q}}k_{\textup{B}}^{-1}T^{-1}Q^{2}}{\left[\left(x_{i}-x_{j}\right)^{2}+\left(y_{i}-y_{j}\right)^{2}+\left(z_{i}-z_{j}\right)^{2}\right]^{1/2}}\right)\nonumber \\
 & \times\exp\left[-\left(\frac{m_{i}g+QE}{k_{\textup{B}}T}\right)z_{i}\right]d^{3}\mathbf{R}_{i}.\label{eq:Ps1}\end{align}

\subsubsection{Approximation}

\begin{figure}
\begin{centering}
\includegraphics[scale=0.45]{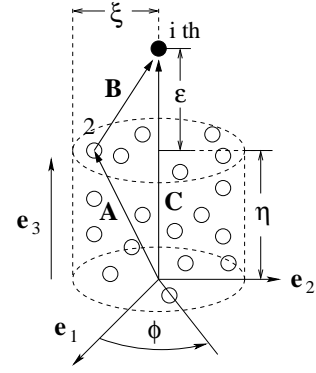} 
\par\end{centering}

\caption{Particles are compressed in $-\mathbf{e}_{3}$ direction. The $i\textup{th}$
particle is the top most particle and all other particles are ahead
of it in the direction of compression. \label{fig:cylinder}}

\end{figure}

The presence of repulsive Coulomb interaction, \[
\sum_{j\neq i}^{N}\frac{-k_{\textup{q}}k_{\textup{B}}^{-1}T^{-1}Q^{2}}{\left[\left(x_{i}-x_{j}\right)^{2}+\left(y_{i}-y_{j}\right)^{2}+\left(z_{i}-z_{j}\right)^{2}\right]^{1/2}},\]
 makes Eq. (\ref{eq:Ps1}) difficult and this term must be approximated.
The configuration depicted in Fig. \ref{fig:cylinder} is referred
for the analysis.

As the $i\textup{th}$ particle gets compressed in the direction of
electric field, it experiences net electric field given by \[
\mathbf{E}_{\textup{net}}=\mathbf{E}_{\textup{el}}+\mathbf{E}_{\textup{rep}},\]
 where $\mathbf{E}_{\textup{el}}$ is the electric field inside the
chamber generated by external electrodes and $\mathbf{E}_{\textup{rep}}$
is the electric field produced by all other particles inside the chamber.
Because $\mathbf{E}_{\textup{el}}$ and $\mathbf{E}_{\textup{rep}}$
are oppositely directed, the magnitude $\mathbf{\mathnormal{E}}_{\textup{net}}\equiv\left\Vert \mathbf{E}_{\textup{net}}\right\Vert $
is given by \begin{equation}
\mathbf{\mathnormal{E}}_{\textup{net}}=\mathnormal{E}_{\textup{el}}-E_{\textup{rep}}.\label{eq:E_net}\end{equation}
 To estimate $E_{\textup{rep}},$ Fig. \ref{fig:cylinder} is considered.
The vectors $\mathbf{A},$ $\mathbf{B},$ and $\mathbf{C}$ satisfy
\begin{equation}
\mathbf{B}=\mathbf{C}-\mathbf{A}.\label{eq:B_C_A}\end{equation}
 In cylindrical coordinates, $\left(\rho,\phi,z\right),$ $\mathbf{A}$
and $\mathbf{C}$ become \begin{align*}
\mathbf{A} & =\rho_{2}\cos\left(\phi_{2}\right)\mathbf{e}_{1}+\rho_{2}\sin\left(\phi_{2}\right)\mathbf{e}_{2}+z_{2}\mathbf{e}_{3},\\
\mathbf{C} & =\left(\eta+\varepsilon\right)\mathbf{e}_{3},\end{align*}
 where $\left(\rho_{2},\phi_{2},z_{2}\right)$ represents the cylindrical
coordinates for particle labeled as $2$ in Fig. \ref{fig:cylinder}.
With $\mathbf{A}$ and $\mathbf{C}$ thus defined, Eq. (\ref{eq:B_C_A})
becomes \[
\mathbf{B}=-\rho_{2}\cos\left(\phi_{2}\right)\mathbf{e}_{1}-\rho_{2}\sin\left(\phi_{2}\right)\mathbf{e}_{2}+\left(\eta+\varepsilon-z_{2}\right)\mathbf{e}_{3}.\]
 At $\mathbf{C},$ the electric field contributed from particle labeled
as $2$ is given by \begin{align*}
\mathbf{E}_{2} & =k_{\textup{q}}Q\frac{\mathbf{B}}{\left\Vert \mathbf{B}\right\Vert ^{3}}\\
 & =\frac{k_{\textup{q}}Q\left[-\rho_{2}\cos\phi_{2}\mathbf{e}_{1}-\rho_{2}\sin\phi_{2}\mathbf{e}_{2}+\left(\eta+\varepsilon-z_{2}\right)\mathbf{e}_{3}\right]}{\left[\rho_{2}^{2}+\left(\eta+\varepsilon-z_{2}\right)^{2}\right]^{3/2}}.\end{align*}
 Because both gravitational and electrical forces are assumed to depend
on $z$ coordinate only, the $\mathbf{e}_{1}$ and $\mathbf{e}_{2}$
components of $\mathbf{E}_{2}$ average to zero to become \[
\mathbf{E}_{2}=\frac{k_{\textup{q}}Q\left(\eta+\varepsilon-z_{2}\right)}{\left[\rho_{2}^{2}+\left(\eta+\varepsilon-z_{2}\right)^{2}\right]^{3/2}}\mathbf{e}_{3}.\]
 All particles in the cylinder, not just the particle labeled as $2$
in Fig. \ref{fig:cylinder}, contributes to form $\mathbf{E}_{\mathrm{\textnormal{rep}}}.$
Hence, \begin{equation}
\mathbf{E}_{\mathrm{\textnormal{rep}}}=k_{\textup{q}}\sum_{j=1}^{N-1}\frac{Q\left(\eta+\varepsilon-z_{j}\right)}{\left[\rho_{j}^{2}+\left(\eta+\varepsilon-z_{j}\right)^{2}\right]^{3/2}}\mathbf{e}_{3},\label{eq:E_total_0}\end{equation}
 where $N$ is the number of charged particles in the chamber. For
$N$ sufficiently large, the coordinates $\rho_{j}$ and $z_{j}$
can be replaced by $\rho_{j}\rightarrow\rho$ and $z_{j}\rightarrow z.$
In the continuum limit, the summation symbol gets replaced by \[
\sum_{j=1}^{N-1}\rightarrow\int_{z=0}^{\eta}\int_{\rho=0}^{\xi}\int_{\phi=0}^{2\pi}\rho d\phi d\rho dz\]
 and the charge $Q$ is replaced by \[
Q\rightarrow\frac{Q_{\mathrm{\textnormal{tot}}}}{\textup{Volume}}=\frac{1}{2\pi\xi\eta}\sum_{j=1}^{N-1}Q=\frac{\left(N-1\right)Q}{2\pi\xi\eta},\]
 where volume is that of cylinder illustrated in Fig. \ref{fig:cylinder}
and $Q_{\mathrm{\textnormal{tot}}}$ is the total charge inside it.
In the continuum limit then, where $N$ is assumed to be sufficiently
large, Eq. (\ref{eq:E_total_0}) can be approximated by \begin{align*}
\mathbf{E}_{\textup{rep}}= & \frac{k_{\textup{q}}\left(N-1\right)Q}{2\pi\xi\eta}\\
 & \times\int_{z=0}^{\eta}\int_{\rho=0}^{\xi}\int_{\phi=0}^{2\pi}\frac{\left(\eta+\varepsilon-z\right)\rho d\phi d\rho dz}{\left[\rho^{2}+\left(\eta+\varepsilon-z\right)^{2}\right]^{3/2}}\mathbf{e}_{3},\end{align*}
 which result, integrating over the $\phi,$ becomes \begin{align}
\mathbf{E}_{\mathrm{\textnormal{rep}}} & =\frac{k_{\textup{q}}\left(N-1\right)Q}{\xi\eta}\int_{z=0}^{\eta}\int_{\rho=0}^{\xi}\frac{\left(\eta+\varepsilon-z\right)\rho d\rho dz}{\left[\rho^{2}+\left(\eta+\varepsilon-z\right)^{2}\right]^{3/2}}\mathbf{e}_{3}.\label{eq:E_total_1}\end{align}
 With the change of variable, \[
x=\rho^{2}+\left(\eta+\varepsilon-z\right)^{2},\quad dx=2\rho d\rho,\]
 the $\rho$ integral in Eq. (\ref{eq:E_total_1}) becomes \begin{align*}
\int\frac{\left(\eta+\varepsilon-z\right)\rho d\rho}{\left[\rho^{2}+\left(\eta+\varepsilon-z\right)^{2}\right]^{3/2}} & \rightarrow\frac{\left(\eta+\varepsilon-z\right)}{2}\int\frac{dx}{x\sqrt{x}}\\
 & =-\frac{\left(\eta+\varepsilon-z\right)}{\sqrt{x}}.\end{align*}
 With $x$ reverted back to the original variable, the $\rho$ integral
becomes \begin{align*}
\int_{\rho=0}^{\xi}\frac{\left(\eta+\varepsilon-z\right)\rho d\rho}{\left[\rho^{2}+\left(\eta+\varepsilon-z\right)^{2}\right]^{3/2}} & =\left.-\frac{\left(\eta+\varepsilon-z\right)}{\sqrt{\rho^{2}+\left(\eta+\varepsilon-z\right)^{2}}}\right|_{0}^{\xi}\\
 & =1-\frac{\left(\eta+\varepsilon-z\right)}{\sqrt{\xi^{2}+\left(\eta+\varepsilon-z\right)^{2}}}.\end{align*}
 Insertion of the result into Eq. (\ref{eq:E_total_1}) yields \begin{align}
\mathbf{E}_{\textup{tot}} & \approx\frac{k_{\textup{q}}\left(N-1\right)Q}{\xi}\mathbf{e}_{3}-\frac{k_{\textup{q}}\left(N-1\right)Q}{\xi\eta}\nonumber \\
 & \times\int_{z=0}^{\eta}\frac{\eta+\varepsilon-z}{\sqrt{\xi^{2}+\left(\eta+\varepsilon-z\right)^{2}}}dz\mathbf{e}_{3}.\label{eq:E_total_2}\end{align}
 With the change of variable, \[
y=\xi^{2}+\left(\eta+\varepsilon-z\right)^{2},\quad dy=-2\left(\eta+\varepsilon-z\right)dz,\]
 the $z$ integral in Eq. (\ref{eq:E_total_2}) becomes \[
\int\frac{\eta+\varepsilon-z}{\sqrt{\xi^{2}+\left(\eta+\varepsilon-z\right)^{2}}}dz\rightarrow-\frac{1}{2}\int\frac{dy}{\sqrt{y}}=-\sqrt{y}.\]
 With $y$ reverted back to the original variable, the $z$ integral
becomes \begin{align*}
\int_{z=0}^{\eta}\frac{\eta+\varepsilon-z}{\sqrt{\xi^{2}+\left(\eta+\varepsilon-z\right)^{2}}}dz & =\left.\vphantom{\frac{\frac{1^{2}}{1}}{\frac{1^{2}}{1}}}-\sqrt{\xi^{2}+\left(\eta+\varepsilon-z\right)^{2}}\right|_{0}^{\eta}\\
 & =\sqrt{\xi^{2}+\left(\eta+\varepsilon\right)^{2}}-\sqrt{\xi^{2}+\varepsilon^{2}}.\end{align*}
 Insertion of the result into Eq. (\ref{eq:E_total_2}) gives the
$\mathbf{E}_{\textup{rep}},$ \begin{align*}
\mathbf{E}_{\textup{rep}} & \approx\frac{k_{\textup{q}}\left(N-1\right)Q}{\xi\eta}\left[\eta-\sqrt{\xi^{2}+\left(\eta+\varepsilon\right)^{2}}+\sqrt{\xi^{2}+\varepsilon^{2}}\right]\mathbf{e}_{3}.\end{align*}
 Since the $z$ coordinate of the $i\textup{th}$ particle is given
by \[
z_{i}=\eta+\varepsilon,\quad\eta=z_{i}-\varepsilon,\]
 the expression for $\mathbf{E}_{\textup{rep}}$ may be rewritten
in terms of $z_{i}$ as \begin{align*}
\mathbf{E}_{\textup{rep}} & \approx k_{\textup{q}}\left(N-1\right)Q\left[\frac{1}{\xi}-\frac{\sqrt{1+z_{i}^{2}/\xi^{2}}}{z_{i}-\varepsilon}+\frac{\sqrt{1+\varepsilon^{2}/\xi^{2}}}{z_{i}-\varepsilon}\right]\mathbf{e}_{3}.\end{align*}
 The parameter $\varepsilon$ has been introduced for a mathematical
convenience to assure that the $i\textup{th}$ particle is the upper
most particle residing at the top surface of the compressed volume.
Taking the limit $\varepsilon\rightarrow0,$ the previous expression
for $\mathbf{E}_{\textup{rep}}$ becomes \begin{equation}
\mathbf{E}_{\textup{rep}}\approx k_{\textup{q}}\left(N-1\right)Q\left(\frac{1}{z_{i}}+\frac{1}{\xi}-\sqrt{\frac{1}{z_{i}^{2}}+\frac{1}{\xi^{2}}}\right)\mathbf{e}_{3}.\label{eq:E_repulsion}\end{equation}
 Since $z_{i}$ is the $z$ coordinate for the $i\textup{th}$ particle,
which is the particle residing at the top surface of the compressed
volume in Fig. \ref{fig:cylinder}, the $\mathbf{E}_{\textup{rep}}$
defined in Eq. (\ref{eq:E_repulsion}) represents the Coulomb repulsion
acting on the particle residing at the top surface of the compressed
volume from all other ones within the compressed volume. Insertion
of Eq. (\ref{eq:E_repulsion}) into Eq. (\ref{eq:E_net}) gives \[
\mathbf{\mathnormal{E}}_{\textup{net}}\approx\mathnormal{E}_{\textup{el}}-k_{\textup{q}}\left(N-1\right)Q\left(\frac{1}{z_{i}}+\frac{1}{\xi}-\sqrt{\frac{1}{z_{i}^{2}}+\frac{1}{\xi^{2}}}\right).\]
 The $\mathbf{\mathnormal{E}}_{\textup{net}}$ in current form is
only an approximation because the expression for $\mathbf{E}_{\textup{rep}},$
Eq. (\ref{eq:E_repulsion}), is an approximation. The equality can
be made by replacing $Q\rightarrow Q_{\textup{eff}},$ where $Q_{\textup{eff}}$
is the effective charge to be determined experimentally. With $Q_{\textup{eff}},$
the expression for $\mathbf{\mathnormal{E}}_{\textup{net}}$ becomes
\begin{equation}
\mathbf{\mathnormal{E}}_{\textup{net}}=E-k_{\textup{q}}\left(N-1\right)Q_{\textup{eff}}\left(\frac{1}{z_{i}}+\frac{1}{\xi}-\sqrt{\frac{1}{z_{i}^{2}}+\frac{1}{\xi^{2}}}\right),\label{eq:E_net_final}\end{equation}
 where the subscript $\textup{el}$ of $\mathnormal{E}_{\textup{el}}$
has been dropped for convenience.

What is the implication of $\mathbf{\mathnormal{E}}_{\textup{net}}$?
The $U_{i}$ of Eq. (\ref{eq:Ui}), which is the energy term assumed
by the $i\textup{th}$ charged particle under the influence of external
forces, can be rearranged in form as \begin{align*}
U_{i}= & \frac{\mathbf{p}_{i}^{2}}{2m_{i}}+m_{i}gz_{i}+Q\left\{ \vphantom{\frac{\frac{\frac{1}{1}}{\frac{1}{1}}}{\frac{\frac{1}{1}}{\frac{1}{1}}}}z_{i}E\right.\\
 & \left.+\sum_{j\neq i}^{N}\frac{k_{\textup{q}}Q}{\left[\left(x_{i}-x_{j}\right)^{2}+\left(y_{i}-y_{j}\right)^{2}+\left(z_{i}-z_{j}\right)^{2}\right]^{1/2}}\vphantom{\frac{\frac{\frac{1}{1}}{\frac{1}{1}}}{\frac{\frac{1}{1}}{\frac{1}{1}}}}\right\} .\end{align*}
 One notices that the term in the summation is the electric field
contribution from the $j\textup{th}$ particle acting on the $i\textup{th}$
particle, \begin{align*}
\left\Vert \mathbf{E}_{\textup{rep},j}\right\Vert  & =\frac{k_{\textup{q}}Q}{\left[\left(x_{i}-x_{j}\right)^{2}+\left(y_{i}-y_{j}\right)^{2}+\left(z_{i}-z_{j}\right)^{2}\right]^{1/2}}.\end{align*}
 Furthermore, one finds \begin{eqnarray*}
E_{\textup{rep}}\equiv\left\Vert \mathbf{E}_{\textup{rep}}\right\Vert =\sum_{j\neq i}^{N}\left\Vert \mathbf{E}_{\textup{rep},j}\right\Vert , &  & E=\left\Vert \mathbf{E}_{\textup{el}}\right\Vert ,\end{eqnarray*}
 and $U_{i}$ can be equivalently expressed as \begin{align*}
U_{i} & =\frac{\mathbf{p}_{i}^{2}}{2m_{i}}+m_{i}gz_{i}+Q\left(z_{i}E+E_{\textup{rep}}\right).\end{align*}
 Since $E_{\textup{rep}}=\mathnormal{E}_{\textup{el}}-\mathbf{\mathnormal{E}}_{\textup{net}},$
Eq. (\ref{eq:E_net}), the $U_{i}$ becomes \begin{align}
U_{i} & =\frac{\mathbf{p}_{i}^{2}}{2m_{i}}+m_{i}gz_{i}+Q\left(z_{i}+1\right)\mathnormal{E}-Q\mathbf{\mathnormal{E}}_{\textup{net}}.\label{eq:Ui_approximation}\end{align}
 Insertion of Eq. (\ref{eq:Ui_approximation}) into Eq. (\ref{eq:Ps0})
gives an alternate expression for probability density for finding
particle with its center of mass position in the ranges $\left(\mathbf{R}_{i};d\mathbf{R}_{i}\right)$
and $\left(\mathbf{p}_{i};d\mathbf{p}_{i}\right),$ \begin{align}
P_{s}\left(\mathbf{R}_{i},\mathbf{p}_{i}\right)d^{3}\mathbf{R}_{i}d^{3}\mathbf{p}_{i}\propto & \exp\left[-\left(\frac{m_{i}g+QE}{k_{\textup{B}}T}\right)z_{i}\right.\nonumber \\
 & \left.+\frac{Q\left(\mathbf{\mathnormal{E}}_{\textup{net}}-E\right)}{k_{\textup{B}}T}\right]d^{3}\mathbf{R}_{i}\nonumber \\
 & \times\exp\left(-\frac{\mathbf{p}_{i}^{2}}{2m_{i}k_{\textup{B}}T}\right)d^{3}\mathbf{p}_{i},\label{eq:Ps20}\end{align}
 which is different from the previous expression, Eq. (\ref{eq:Ps1}),
but now manageable. With Eq. (\ref{eq:E_net_final}) inserted for
$\mathbf{\mathnormal{E}}_{\textup{net}},$ Eq. (\ref{eq:Ps20}) becomes
\begin{align}
 & P_{s}\left(\mathbf{R}_{i},\mathbf{p}_{i}\right)d^{3}\mathbf{R}_{i}d^{3}\mathbf{p}_{i}\nonumber \\
 & \propto\exp\left[\vphantom{\frac{\frac{\frac{1}{1}}{1}}{\frac{\frac{1}{1}}{\frac{1}{1}}}}\frac{k_{\textup{q}}\left(1-N\right)QQ_{\textup{eff}}}{k_{\textup{B}}T}\left(\frac{1}{z_{i}}+\frac{1}{\xi}-\sqrt{\frac{1}{z_{i}^{2}}+\frac{1}{\xi^{2}}}\right)\right.\nonumber \\
 & \left.-\left(\frac{m_{i}g+QE}{k_{\textup{B}}T}\right)z_{i}\vphantom{\frac{\frac{\frac{1}{1}}{1}}{\frac{\frac{1}{1}}{\frac{1}{1}}}}\right]\exp\left(-\frac{\mathbf{p}_{i}^{2}}{2m_{i}k_{\textup{B}}T}\right)d^{3}\mathbf{R}_{i}d^{3}\mathbf{p}_{i},\label{eq:Ps21}\end{align}
 Equation (\ref{eq:Ps21}) may be integrated over all possible $x$
and $y$ values lying within in the container and each components
of the momentum may be integrated from $-\infty$ to $\infty,$ \begin{align}
P_{s}\left(z_{i}\right)dz_{i}\propto & \int_{p_{i\textup{x}}}\int_{p_{i\textup{y}}}\int_{p_{i\textup{\textup{z}}}}\exp\left(-\frac{p_{i\textup{x}}^{2}+p_{i\textup{y}}^{2}+p_{i\textup{z}}^{2}}{2m_{i}k_{\textup{B}}T}\right)\nonumber \\
 & \times dp_{i\textup{x}}dp_{i\textup{y}}dp_{i\textup{z}}\exp\left[\vphantom{\frac{\frac{\frac{1}{1}}{1}}{\frac{\frac{1}{1}}{\frac{1}{1}}}}\frac{k_{\textup{q}}\left(1-N\right)QQ_{\textup{eff}}}{k_{\textup{B}}T}\right.\nonumber \\
 & \times\left(\frac{1}{z_{i}}+\frac{1}{\xi}-\sqrt{\frac{1}{z_{i}^{2}}+\frac{1}{\xi^{2}}}\right)\nonumber \\
 & \left.-\left(\frac{m_{i}g+QE}{k_{\textup{B}}T}\right)z_{i}\vphantom{\frac{\frac{\frac{1}{1}}{1}}{\frac{\frac{1}{1}}{\frac{1}{1}}}}\right]dz_{i}\int_{x}\int_{y}dx_{i}dy_{i}.\label{eq:Ps22}\end{align}
 The double integral over $x$ and $y$ gives slice area of the chamber
at $z=z^{\prime},$ $0<z^{\prime}<L_{\textup{z}},$ \begin{equation}
\int_{x}\int_{y}dx_{i}dy_{i}=\pi\xi^{2},\label{eq:Ps22a}\end{equation}
 where $\xi$ is the radius of cylindrical chamber depicted in Fig.
\ref{fig:cylinder}. The momentum integrals are obtained utilizing
the well known integral formula\cite{Reif} \[
I\left(n\right)\equiv\int_{0}^{\infty}x^{n}\exp\left(-\alpha x^{2}\right)dx,\quad n\geq0,\]
 where solutions are given by \[
I\left(0\right)=\frac{1}{2}\sqrt{\frac{\pi}{\alpha}},\quad I\left(1\right)=\frac{1}{2\alpha},\quad I\left(2\right)=\frac{1}{4\alpha}\sqrt{\frac{\pi}{\alpha}},\quad\textup{etc.}\]
 The momentum integrals become \begin{align}
 & \int_{p_{i\textup{x}}}\int_{p_{i\textup{y}}}\int_{p_{i\textup{\textup{z}}}}\exp\left(-\frac{p_{i\textup{x}}^{2}+p_{i\textup{y}}^{2}+p_{i\textup{z}}^{2}}{2m_{i}k_{\textup{B}}T}\right)dp_{i\textup{x}}dp_{i\textup{y}}dp_{i\textup{z}}\nonumber \\
 & \qquad\qquad=\frac{1}{8}\left(2\pi m_{i}k_{\textup{B}}T\right)^{\frac{3}{2}}.\label{eq:Ps22b}\end{align}
\begin{figure}
\begin{centering}
\includegraphics[scale=0.45]{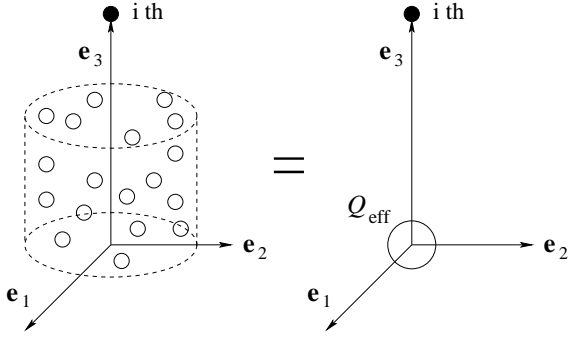} 
\par\end{centering}

\caption{Transformation to the two body problem by introduction of effective
charge for the particles inside the imaginary cylinder. \label{fig:cylinder_Qeff}}

\end{figure}
 With Eqs. (\ref{eq:Ps22a}) and (\ref{eq:Ps22b}), Eq. (\ref{eq:Ps22})
becomes \begin{align*}
P_{s}\left(z_{i}\right)dz_{i}\propto & \frac{1}{8}\pi\xi^{2}\left(2\pi m_{i}k_{\textup{B}}T\right)^{\frac{3}{2}}\exp\left[\vphantom{\frac{\frac{\frac{1}{1}}{1}}{\frac{\frac{1}{1}}{\frac{1}{1}}}}\frac{k_{\textup{q}}\left(1-N\right)QQ_{\textup{eff}}}{k_{\textup{B}}T}\right.\\
 & \times\left(\frac{1}{z_{i}}+\frac{1}{\xi}-\sqrt{\frac{1}{z_{i}^{2}}+\frac{1}{\xi^{2}}}\right)\\
 & \left.-\left(\frac{m_{i}g+QE}{k_{\textup{B}}T}\right)z_{i}\vphantom{\frac{\frac{\frac{1}{1}}{1}}{\frac{\frac{1}{1}}{\frac{1}{1}}}}\right]dz_{i}.\end{align*}
 With the following definitions, \begin{eqnarray}
\alpha=\frac{k_{\textup{q}}\left(1-N\right)QQ_{\textup{eff}}}{k_{\textup{B}}T}, &  & \beta=\frac{m_{i}g+QE}{k_{\textup{B}}T},\label{eq:alpha_beta}\end{eqnarray}
 the previous expression for $P_{s}\left(z_{i}\right)dz_{i}$ simplifies
to \begin{align}
P_{s}\left(z_{i}\right)dz_{i}= & \frac{1}{8}C\pi\xi^{2}\left(2\pi m_{i}k_{\textup{B}}T\right)^{\frac{3}{2}}\nonumber \\
 & \times\exp\left[\alpha\left(\frac{1}{z_{i}}+\frac{1}{\xi}-\sqrt{\frac{1}{z_{i}^{2}}+\frac{1}{\xi^{2}}}\right)-\beta z_{i}\right]dz_{i},\label{eq:Ps23}\end{align}
 where $C$ is the constant of proportionality to be determined from
the normalization condition, $\int_{0}^{L_{\textup{z}}}P_{s}\left(z_{i}\right)dz_{i}=1.$
It can be shown \begin{align}
C= & \frac{8}{\pi\xi^{2}}\left(2\pi m_{i}k_{\textup{B}}T\right)^{-\frac{3}{2}}\left\{ \int_{0}^{L_{\textup{z}}}\exp\left[\alpha\left(\vphantom{\sqrt{\frac{1}{x_{i}^{2}}}}\frac{1}{x_{i}}+\frac{1}{\xi}\right.\right.\right.\nonumber \\
 & \left.\left.\left.-\sqrt{\frac{1}{x_{i}^{2}}+\frac{1}{\xi^{2}}}\right)-\beta x_{i}\right]dx_{i}\right\} ^{-1},\label{eq:CCCCC}\end{align}
 where $x_{i}$ is a dummy integration variable and it should not
be confused with the coordinate $x$ of the cylinder. 

What can be said about $Q_{\textup{eff}}$ defined in $\alpha$? The
effective Coulomb repulsion from the remaining $N-1$ charged particles
inside the chamber acting on the $i\textup{th}$ charged particle,
see Fig. \ref{fig:cylinder}, is proportional to \[
\mathcal{F}_{\textup{N-1}}\propto\left(N-1\right)Q_{\textup{eff}},\]
 where $Q_{\textup{eff}}$ must be determined empirically from measurements.
In principle, $Q_{\textup{eff}}$ takes into account the spatial configuration
of the $N-1$ charged particles in the system because it effectively
describes the system, which is illustrated in Fig. \ref{fig:cylinder},
in terms of the two body problem (see Fig. \ref{fig:cylinder_Qeff}).
Because the total charge in the imaginary cylinder must be conserved,
it must be true that \[
0<\left(N-1\right)Q_{\textup{eff}}\leq\left(N-1\right)Q.\]
 And, this implies the condition \begin{equation}
0<Q_{\textup{eff}}\leq Q.\label{eq:Qeff_condition}\end{equation}
 For describing the trend of volume compression involving charged
particles, Eq. (\ref{eq:Qeff_condition}) provides the way to estimate
$Q_{\textup{eff}}.$ Once $Q_{\textup{eff}}$ is defined, Eq. (\ref{eq:Ps23})
may be plotted for the most probable height of the compressed volume,
which volume contains the $N$ charged particles in the system. That
being said, combining Eqs. (\ref{eq:Ps23}) and (\ref{eq:CCCCC}),
the probability density for the most probable height of the compressed
volume containing $N$ charged particles becomes \begin{align}
P_{s}\left(z_{i}\right)dz_{i}= & \left\{ \int_{0}^{L_{\textup{z}}}\exp\left[\alpha\left(\frac{1}{x_{i}}+\frac{1}{\xi}-\sqrt{\frac{1}{x_{i}^{2}}+\frac{1}{\xi^{2}}}\right)\right.\right.\nonumber \\
 & \left.\left.-\beta x_{i}\vphantom{\sqrt{\frac{1}{x_{i}^{2}}}}\right]dx_{i}\right\} ^{-1}\exp\left[\alpha\left(\frac{1}{z_{i}}+\frac{1}{\xi}-\sqrt{\frac{1}{z_{i}^{2}}+\frac{1}{\xi^{2}}}\right)\right.\nonumber \\
 & \left.-\beta z_{i}\vphantom{\sqrt{\frac{1}{z_{i}^{2}}}}\right]dz_{i},\label{eq:Ps24}\end{align}
 where $\alpha$ and $\beta$ are defined in Eq. (\ref{eq:alpha_beta}).

\subsubsection{Result}

Before plotting $P_{s}\left(z_{i}\right),$ I shall explicitly define
the charge $Q,$ particle mass $m,$ and the electric field magnitude
$E.$

In nature, the charge is quantized and, therefore, it is convenient
to express $Q$ in terms of the charge number $n_{\textup{e}},$ \begin{equation}
Q_{i}=n_{\textup{e}}q,\quad q=1.602\times10^{-19}\textup{ C},\label{eq:Q-DEF}\end{equation}
 where $q$ is the fundamental charge unit and $n_{\textup{e}}=0,1,2,3,\cdots,$
is the number of electrons removed from the particle.

For the sake of simple analysis, the particles in the system are assumed
to be spheres of identical radius. The mass of each particle would
then be given by \begin{equation}
m=\frac{4}{3}\pi r^{3}\rho_{\textup{m}},\label{eq:m-DEF}\end{equation}
 where $r$ is the radius of sphere and $\rho_{\textup{m}}$ is the
mass density.

Finally, the electric field generated in the chamber by control electrodes
is \begin{equation}
E\approx\frac{V_{0}}{L_{\textup{e}}},\label{eq:E}\end{equation}
 where $V_{0}$ is the voltage applied to the top electrode (the other
electrode has been grounded). The approximation $\approx$ in electric
field comes about because the effects of passivation layers inside
the chamber have been neglected for simplicity.

\begin{figure}
\begin{centering}
\includegraphics[scale=0.45]{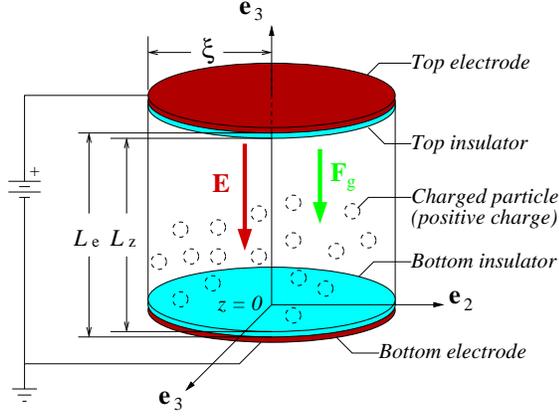} 
\par\end{centering}

\caption{(Color online) Configuration used to plot $P_{s}\left(z_{i}\right)$
defined in Eq. (\ref{eq:Ps24}). \label{fig:cylinder2}}

\end{figure}

\begin{figure}[H]
\begin{centering}
\includegraphics[scale=0.65]{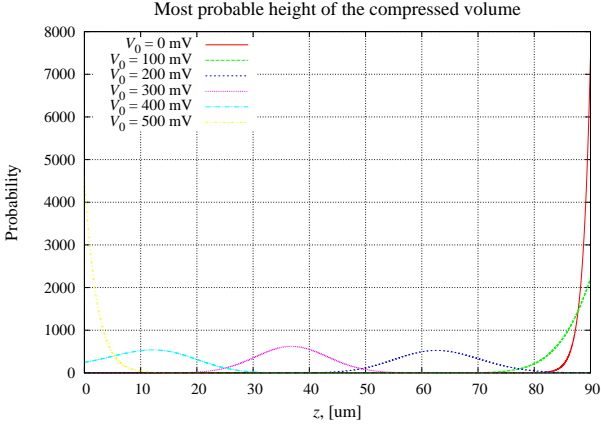} 
\par\end{centering}

\caption{\label{fig:probability} (Color online) Most probable height of the
compressed volume containing charged particles.}

\end{figure}

\begin{figure}[H]
\begin{centering}
\includegraphics[scale=0.65]{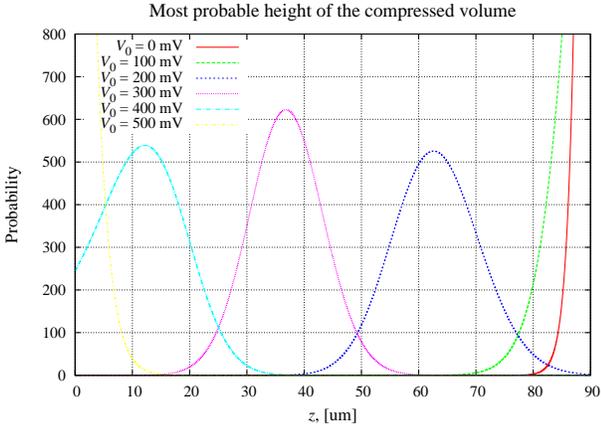} 
\par\end{centering}

\caption{\label{fig:probability_zoom} (Color online) The three small peaks
in Fig. \ref{fig:probability} are magnified for detail. Each peak
represents the most probable height for the compressed volume, where
there are $N$ charged nanoparticles inside the volume. }

\end{figure}

Having defined $Q,$ $m,$ and $\mathbf{E},$ the configuration illustrated
in Fig. \ref{fig:cylinder2} is referred to plot $P_{s}\left(z_{i}\right),$
Eq. (\ref{eq:Ps24}). The parameters for the configuration are given
the following values:

\[
\left\{ \begin{array}{c}
L_{\textup{e}}=100\,\mu\textup{m},\; L_{\textup{z}}=90\,\mu\textup{m},\;\xi=50\,\mu\textup{m},\\
\rho_{\textup{m}}=2.7\,\textup{g}\,\textup{cm}^{-3},\; r=50\,\textup{nm},\\
T=42\,^{o}\textup{C},\; N=1000,\; n_{\textup{e}}=15.\end{array}\right.\]
 For the purpose of plotting $P_{s}\left(z_{i}\right)$ defined in
Eq. (\ref{eq:Ps24}), I shall assume, see Eq. (\ref{eq:Qeff_condition}),
\[
Q_{\textup{eff}}=0.97Q,\]
 where $Q$ is defined in Eq. (\ref{eq:Q-DEF}). The voltages of $V_{0}=0\,\textup{V},$
$0.1\,\textup{V},$ $0.2\,\textup{V},$ $0.3\,\textup{V},$ and $0.5\,\textup{V}$
are considered for the top electrode (the bottom electrode is grounded).
With $V_{0}$ thus defined, the electric field generated inside chamber
is given by Eq. (\ref{eq:E}). The gravity of $g=9.8\,\textup{m}/\textup{s}^{2}\geq0$
was assumed inside the chamber. The directions for the electric field
$\mathbf{E}$ and the gravitational force $\mathbf{F}_{\textup{g}}$
are determined from Eqs. (\ref{eq:g_direction}) and (\ref{eq:E_direction}).
Since both $E$ and $g$ are positive, according to the convention
defined in Eqs. (\ref{eq:g_direction}) and (\ref{eq:E_direction}),
the electric field $\mathbf{E}$ and the gravitational force $\mathbf{F}_{\textup{g}}$
are both directed in $-\mathbf{e}_{3},$ which is the negative $z$
axis. The $P_{s}\left(z_{i}\right)$ of Eq. (\ref{eq:Ps24}) is computed
numerically utilizing Simpson method for the integral.\cite{thomas-finney}
The Simpson method routine was coded in FORTRAN 90. That being said,
the results are summarized in Figs. \ref{fig:probability} and \ref{fig:probability_zoom},
where the three smaller peaks of Fig. \ref{fig:probability} are magnified
and replotted in Fig. \ref{fig:probability_zoom}.

At $V_{0}=0\,\textup{V},$ that is, when there is no electric field
inside the chamber other than the static fields from particles, the
particles are distributed to occupy the entire volume of the chamber.
This is indicated by the peak occurring at the physical height of
the chamber, $h_{\textup{c}}=z=L_{\textup{z}}=90\,\mu\textup{m}.$

At $V_{0}=0.2\,\textup{V},$ an electric field of roughly $E\approx2000\,\textup{V}\,\textup{m}^{-1}$
is generated inside the chamber. Because particles are positively
charged, they are compressed in the direction of electric field, i.e.,
the $-z$ direction. This force, which induced particle volume compression,
eventually gets counter balanced by the Coulomb repulsion and the
compression ceases. For the case where control electrode is held at
$V_{0}=0.2\,\textup{V},$ the compression ceases at roughly $h_{\textup{c}}=z\approx63\,\mu\textup{m}$
(see Fig. \ref{fig:probability_zoom}) and this marks the most probable
height of the compressed volume for the case. Finally, with $V_{0}=0.5\,\textup{V}$
applied to the control electrode, the compressed volume state is reached
where all particles are cluttered near the floor of the chamber, thereby
resulting in very high particle density, as illustrated in Fig. \ref{fig:probability}.

\begin{figure}[H]
\begin{centering}
\includegraphics[scale=0.65]{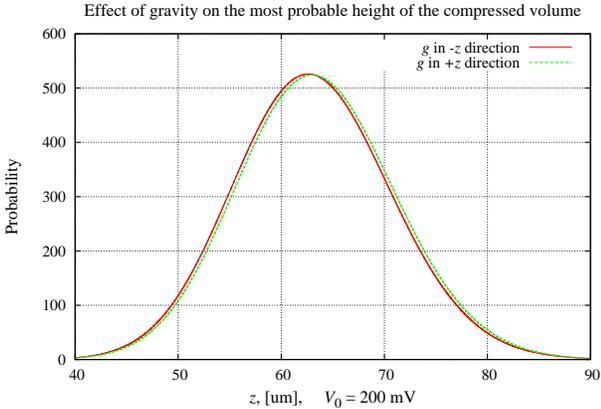} 
\par\end{centering}

\caption{\label{fig:probability_g_effect} (Color online) Gravity has negligible
effect on most probable height of the compressed volume. }

\end{figure}

If charged particles are to be useful for any display applications,
the charged particle system must be insensitive to gravitational effects,
if not negligible. The effect of gravity on the most probable height
for the compressed volume has been investigated by reversing the direction
of gravity (but, keeping all other conditions unchanged) in Fig. \ref{fig:probability_zoom}.
The case where $V_{0}=0.2\,\textup{V}$ was selected for comparison.
The result is shown in Fig. \ref{fig:probability_g_effect}, where
it shows that the most probable height for the compressed volume is
only negligibly affected by the gravity.

\begin{figure}[H]

\begin{centering}
\includegraphics[scale=0.65]{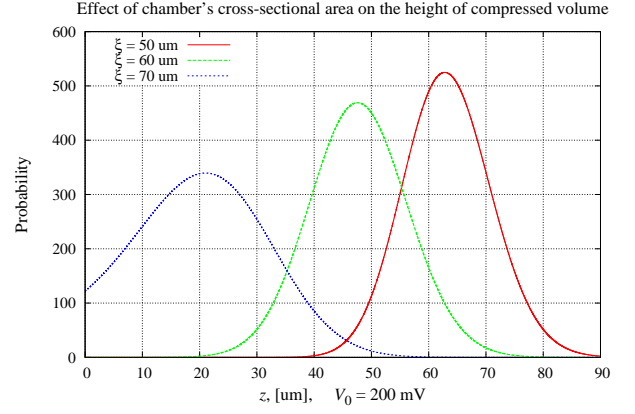} 
\par\end{centering}

\caption{\label{fig:probability_eta_effect} (Color online) The height of the
compressed volume decreases with the increased radius of the cylindrical
chamber.}

\end{figure}

The influence of cylinder radius $\xi$ on the most probable height
for the compressed volume has also been investigated. Again, the case
of $V_{0}=0.2\,\textup{V}$ was selected from Fig. \ref{fig:probability_zoom}
for comparison by considering $\xi=50,$ $60,$ and $70\,\mu\textup{m}.$
All other conditions were kept unmodified. The result, Fig. \ref{fig:probability_eta_effect},
reveals a decrease in height for the most probable compressed volume
with increasing $\xi$ as expected.

\subsection{Transmission intensity}

The intensity of light transmitted through a medium filled with charged
particles goes like, Eq. (\ref{eq:transmitted_intensity}), \begin{eqnarray}
I=I_{0}\exp\left(-\frac{C}{h_{\textup{c}}}\right), &  & C=\frac{36n\pi khNv_{\textup{p}}}{\lambda\left(n^{2}+2\right)^{2}A},\label{eq:I_hc}\end{eqnarray}
 where both $C$ and $h_{\textup{c}}$ have unit of meter, and $h_{\textup{c}}$
is the height of the compressed volume containing charged particles.
Referring to Fig. \ref{fig:probability_zoom}, the compressed height
$h_{\textup{c}}$ corresponds to the $z$ axis where the probability
curve is maximum. In principle, the compression height $h_{\textup{c}}$
can be varied continuously by controlling the voltages applied to
the control electrode. This implies the intensity of light output
from the proposed optical shutter can be varied continuously, thereby
generating continuous grayscale levels for the device. In reality,
the number of grayscale levels that can be achieved in the presented
optical shutter is given by Eq. (\ref{eq:Ngray}), \begin{align*}
N_{\textup{gray}} & =\frac{h-\delta}{\triangle h_{\textup{c}}},\end{align*}
 where the fineness of $\triangle h_{\textup{c}}$ is limited by the
system design.

This work is not the first kind to address potential applications
with charged particles. Szirmai\cite{Szirmai} experimented with alumina
powders to study electrosuspension as early as 1990's. In Szirmai's\cite{Szirmai}
experiment, alumina powder of $3\,\mu\textup{m}$ in diameter was
placed inside an electrically insulating cylindrical vessel, which
is similar in configuration with Fig. \ref{fig:cylinder2}. The initial
charging of alumina powder was done by a process of field emission,
which can be achieved by applying high voltage to the control electrodes(field
emission is the phenomenon in which electrons get emitted from the
surface of host material, such as nanoparticles, due to the presence
of high electric fields). Szirmai,\cite{Szirmai} however, does not
quantitatively address the compression states of volume containing
charged particles in terms of the design parameters, as his motive
was not in discussing possible applications of charged particles for
displays. In his experiment, the control electrodes, in principle,
could be supplied with whatever high voltages required by it to do
the job; therefore, the quantitative understanding of how design parameters,
such as $L_{\textup{e}},$ $L_{\textup{z}},$ $\xi,$ $\rho_{\textup{m}},$
$r,$ $T,$ $N,$ $n_{\textup{e}},$ $V_{0},$ and $Q_{\textup{eff}}\propto Q,$
enter into the picture of particle volume compression never was an
issue.

The competition has always been fierce and it will always remain so
among different display manufacturers. In the near future, when paperlike
displays become dominant, the most important deciding factor to who
stays in and goes out of business would be determined by the power
efficiency of their products. That being said, a low operation voltage
for the control electrodes is crucial for all E-paper technologies
and the proposed device based on charged particles is no exception.
The light intensity out of each sub-pixel based on proposed charged
particle display technology varies as illustrated in Eq. (\ref{eq:I_hc}),
where $h_{\textup{c}}$ is the most probable compression height corresponding
to the voltage difference of $V_{0}$ applied to the control electrodes,
see Fig. \ref{fig:probability}. With the $V_{0}$ restricted to certain
range, say $0\,\textup{V}\leq V_{0}\leq1\,\textup{V},$ one cannot
arbitrarily choose the other parameters which constitute the design
parameters, i.e., $L_{\textup{e}},$ $L_{\textup{z}},$ $\xi,$ $\rho_{\textup{m}},$
$r,$ $T,$ $N,$ $n_{\textup{e}},$ and $Q.$ For example, if too
many electrons are removed from each of the aluminum particles, i.e.,
$n_{\textup{e}},$ the voltage of $V_{0}=1\,\textup{V}$ applied to
one of the control electrodes (the other grounded) may not be sufficient
enough to overcome the Coulomb repulsion between particles and compress
the particle volume to a level where dark state is reached, assuming
$V_{0}=0\,\textup{V}$ defines the brightest state. On the other end,
if too many charged particles are present in a chamber, i.e., the
particle number $N,$ the brightest state achieved by setting $V_{0}=0\,\textup{V}$
for the control electrode may be too dark. The quantitative description
of the height $h_{\textup{c}}$ of the compressed particle volume
in terms of the so called {}``design parameters'' thru the expression
$P_{s}\left(z_{i}\right),$ Eq. (\ref{eq:Ps24}), ables the design
of particle based display with potential to generate continuous grayscale.

\subsection{Bistability}

\begin{figure}[H]
\begin{centering}
\includegraphics[scale=0.5]{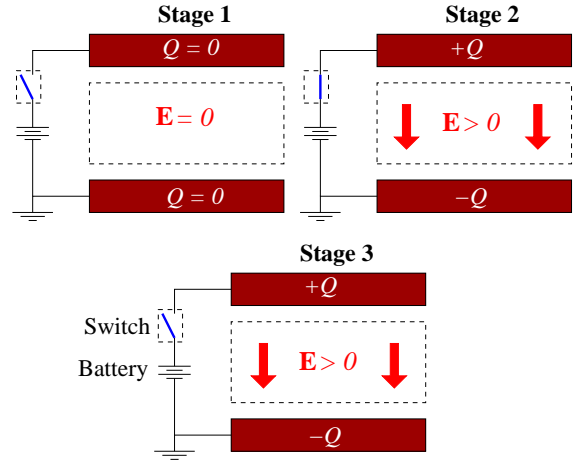} 
\par\end{centering}

\caption{\label{fig:capacitor} (Color online) The bistability is maintained
by the electric field stored in the capacitor.}

\end{figure}

The presented optical shutter based on charged particles portrays
bistability at all states, including the gray states. This is possible
because the two optically transparent electrodes act as a capacitor,
which has the property of sustaining electric fields even when the
device is removed of the power supply. To illustrated how the bistability
is achieved for all states, including the gray states, the illustration
shown in Fig. \ref{fig:capacitor} is considered. I shall begin with
an isolated capacitor, in which the two electrodes of the capacitor
are electrically neutral, resulting in zero electric field inside
the region between the two electrodes. With the switch closed, the
top electrode is quickly accumulated with a net positive charge, $+Q,$
and the bottom electrode gets accumulated with a net negative charge,
$-Q.$ The potential difference between the two electrodes results
in the creation of electric field inside the capacitor, as illustrated
in stage 2 of Fig. \ref{fig:capacitor}. Assuming the charged particles
reside in the region between the two electrodes, the electric field
generated inside the capacitor is responsible for the compression
of volume containing charged particles. Now, when the switch is opened,
the net charge of $+Q$ remains in the top electrode and the net charge
of $-Q$ remains in the bottom electrode, provided the capacitor is
ideal, i.e., free from the leakage of electrical current. Therefore,
for an ideal capacitor, the electric field is maintained forever inside
the region between the two electrodes, thereby sustaining the gray
states even when the device is removed of the power supply.

In the real system, the role of switch is played by a semiconductor
transistor, which is very far from being an ideal switch and it has
finite leakage of current. This deficiency in semiconductor transistor
makes the proposed device only semi-bistable, meaning the device must
be refreshed regularly. This, however, is about to change with the
current developments in micro electromechanical systems (MEMS) based
switches, which literally has zero leakage current for the open state.\cite{MEMS-switch}
Initially, the MEMS based switch has been developed in an attempt
to replace the dynamic random access memory (DRAM) architecture for
the memory sector of business. However, as it lacks in switching speed,
it will be a while before MEMS based switches can permanently replace
the DRAMs. As for its use as a switch in display technology, the MEMS
based switches already show plenty of speed. Combined with MEMS based
switches, which has zero leakage current for the open switch mode,
the proposed optical shutter based on charged particles opens up the
possibility of realizing the bistability mode for all states, including
the grayscale states.

\section{Concluding Remarks}

The pioneering work by Szirmai,\cite{Szirmai} Hattori et al.,\cite{QR-LPD2,QR-LPD3,QR_LPD4}
and others have exposed the potential applications with charged particles.
Utilizing charged particles in display technologies, however, requires
a quantitative understanding of how design parameters, such as $L_{\textup{e}},$
$L_{\textup{z}},$ $\xi,$ $\rho_{\textup{m}},$ $r,$ $T,$ $N,$
$n_{\textup{e}},$ $V_{0},$ and $Q_{\textup{eff}}\propto Q,$ enter
into particle volume compression. In this work, an expression for
the compressed state, which incorporates the design parameters, has
been presented. The result should find its role in the development
of displays based on charged particles.

\section{Acknowledgments}

The author acknowledges the support for this work provided by Samsung
Electronics, Ltd.

\end{document}